# Multiterminal Nanowire Junctions of Silicon: A Theoretical Prediction of Atomic Structure and Electronic Properties

*Pavel V. Avramov*[*, †, ‡], *Leonid A. Chernozatonskii*[#], *Pavel B. Sorokin*[‡, #], *and Mark S. Gordon*[$]

Takasaki-branch, Advanced Science Research Center, Japan Atomic Energy Agency, Takasaki, 370-1292, Japan, L.V. Kirensky Institute of Physics SB RAS, 660036 Krasnoyarsk, Russian Federation, N.M. Emanuel Institute of Biochemical Physics of RAS, 119334 Moscow, Russian Federation, Ames National Laboratory/Department of Chemistry, Iowa State University, Ames, Iowa 50011, USA

## ABSTRACT

Using empirical scheme, atomic structure of a new exotic class of silicon nanoclusters was elaborated upon the central icosahedral core (Si-IC) and pentagonal petals (Si-PP) growing from Si-IC vertexes. It was shown that Si-IC/Si-PP interface formation is energetically preferable. Some experimental observations of silicon nanostructures can be explained by presence of the proposed objects. The Extended Huckel Theory electronic structure calculations demonstrate an ability of the proposed objects to act as nanoscale tunnel junctions.

---

[*] Corresponding author. E.mail: avramov.pavel@jaea.go.jp; Telephone: (81)-27-346-9670; Fax: (81)-27-346-9696.

[†]Advanced Science Research Center.

At present a number of silicon quantum dots (QD) and nanowires (NW) (mostly covered by hydrogen or embedded into silica environment, [1-7]) with polycrystalline structure [1, 2, 4] have been synthesized. The photoelectron experiment [6] directly demonstrates a sharp nature of the Si/SiO$_2$ interface. All silicon nanostructures with saturated surface dangling bonds display pronounced semiconducting properties. The surface tension caused by different type of interfaces or by the formation of surface dimers [8] on the unsaturated Si(100) surfaces closes the band gap of the nanocrystalline (*nc*) silicon. [5] The DFT electronic structure calculations of the NWs [9-11] not covered by a hydrogen or oxidized layer demonstrate the metallic nature of the electronic structure of the objects, whereas saturation of the dangling bonds by hydrogen [12, 13] or by oxidized layer [11] opens semiconducting band gaps different in width and type.

The most realistic atomic models of the small sized NWs and QDs were designed [12, 14] based on the combination of silicon triangular prisms or tetrahedrons with two or four equivalent <111> facets. Combination of 4, 5 or 6 prisms gives square, pentagonal or hexagonal NWs, [14] whereas combination of 20 tetrahedrons produces the icosahedral QDs (IQDs). [12] For $d \leq 5$ *nm*, the structures with pentagonal symmetry (pentagonal NWs or PNWs and IQDs) are energetically preferable among the different types of *nc*-Si. [12, 14, 15] The IQD surface is formed by 20 <111> facets with a Si$_{20}$ dodecahedron in the center. [12] The 12 IQD vertexes are formed by the silicon pentagons. [12]

The pentagonal vertexes are specific points on the icosahedron surface with the maximum curvature. They can serve as the natural starting points for the growth of the PNWs. The (100) surface of each PNW can be rearranged by the formation of the dimer rows parallel (with decreasing of the surface, energetically preferable) or perpendicular (with increasing of the surface) to the main axis of the nanowire. [11, 14]

All PNWs [14] have a central pentagonal prism as the basis, surrounded by several layers of the hexagonal prisms (see Fig. 1a). The PNWs can be classified by the number of prism layers surrounding the central pentagonal prism. The SiPNW(1) [15] corresponds to the central pentagonal

---

‡L.V. Kirensky Institute of Physics

#N.M. Emanuel Institute of Biochemical Physics.

core (the first and the smallest circle surrounding the pentagon in the center of Fig. 1a). The SiPNW(2) (the second circle, Fig. 1a) can be obtained by surrounding the central pentagonal prism with the first layer of hexagonal prisms. In this work we will use the SiPNW(2)s as the building blocks of the complex silicon nanostructures described later.

Like carbon nanotubes, the SiPNW($l$) tips can be covered by caps formed using half of the corresponding IQD($l$)'s (where $l$ is a number of hexagonal layers introduced above),[11] with five <111> triangles formed by cutting each PNW's prisms from one side along the [110] directions. The connection of the SiPNW($l$) with an IQD's vertex can be made through a cavity in the opposite side of the SiNW($l$) with five <111> surfaces obtained by truncation of the prisms along the same [110] directions (Fig. 1b). Later we will call the SiPNW with an IQD cap at one end and a cavity at another one as a Pentagonal Petal (PP).

The combination of the different numbers of PPs (from 1 to 12) with a single IQD produces a set of perfect silicon nanostructures which look like flowers (SiNFs) or stars (Fig. 2). The number of PPs is limited not only by the number of IQD's vertexes (12), but also by the ratio of the number of shells ($l$ and $m$) of corresponding PP($l$)s and the central IQD($m$).

Some exotic snowflake silicon micro-[16] and nano-[17] structures were observed in the experiment. The proposed nanoflowers could serve precursors in formation of the structures.[16,17] For example, the 5-petaled "flower" in the center of Fig.1 [17] has a very similar structure to Fig.2c.

A combination of metallic [9-11] PPs with semiconducting ones, with different band gap widths and types of conductivity, around a central IQD in one SiNF can serve as a background in the developing of a large variety of nanoelectronic devices. To study the electronic structure of such complex SiNFs, we designed and calculated a set (Fig. 3) of SiNFs based on the IQD(2) (100 silicon atoms in total [12]) and 3 PP(2)s. In the case of the pristine SiNF (640 silicon atoms, Fig. 3a) their (100) facets allow the surface silicon atoms to form dimer rows parallel to the PP's main axis with the decreasing of the NW's surface [11,14] (we will call such structures SiNF/D, where D denotes dimers). Covering the surface of the corresponding PPs by hydrogen produces the SiNF/H structure (Fig. 3b, $Si_{640}H_{420}$).

The structure of the surface oxidized layer (Figs. 2a, 2b, 3c and 3d) is more complex. [11] the silicon atoms on the Si (100) surface are connected with each other by bridged oxygen atoms, whereas the edges of the PPs are covered by $SiO_4$ fragments. It allows us to keep natural four- and two-fold coordination of each silicon and oxygen atom respectively and bond the neighboring silicon prisms with each other. Some of the silicon atoms (on the tips of the PPs and on the central IQD) in such a structure cannot be bonded with each other by a bridged oxygen; to keep the tetravalent nature of the silicon, the surface dangling bonds were saturated by OH groups. In general, the SiNF/OOH (Fig. 3c) structure has $Si_{640}/Si_{138}O_{505}H_{38}$ formula. Finally, the SiNF structure with 3 different types of PPs (one metallic PP with surface dimers and two semiconducting ones with dangling bonds saturated by hydrogen and oxidized layer, SiNF/D/H/OOH) was designed (Fig. 3d).

The atomic structure of the objects was optimized using model MM+ potential. [18] Previously the MM+ have been successfully used in numerous studies of silica [19, 20] systems. To study the stability of the systems we calculated the structures of SiPNW(2), SiIQD(2) and SiIQD(4) (600 atoms) and even a $Si_{552}$ cluster of the bulk silicon. The relative stability of the set of SiIQDs and SiPNW(2)s, in respect of the bulk silicon qualitatively confirms the DFT data. [12] Due to the surface tension (Fig. 1c), the longer the SiPNW(2) the higher the relative energy of the system. The icosahedral IQDs have significantly lower energy, with opposite dependence of the relative energy per atom upon the size/number of atoms in the system due to the decreasing of the relative surface.

To study the SiIQD/PP interface we performed calculations of one-petal system SiIQD(2)/PP(2) with $Si_{280}$ formula. The combination of the IQD and PP parts is energetically preferable due to the significant decreasing of the relative binding energy from 0.176 eV/atom ($Si_{295}NW$) and 0.173 eV/atom ($Si_{205}NW$) to 0.160 eV/atom for the SiIQD(2)/PP(2) (Fig. 1c) in respect with the MM+ binding energy of the bulk silicon (0.0 eV/atom). Many experimental STM images (see works of Refs. 3, 4, 7) exhibit the same NW/QD structures.

The structural tension of all SiNF objects affects the atomic structure of the species depending on the exact location of the atoms. The bulk silicon Si-Si distance at the MM+ level ($Si_{512}$ cluster) is equal to 2.216 Å, whereas the Si-Si bond of the IQD's central dodecahedron is equal to 2.303 Å. The Si-Si distance in the SiIQD's second sphere is closer to the bulk one (2.289 Å). The Si-Si distances of the PP part (2.303 ÷ 2.357 Å) deviate bigger from the bulk one. The deviation of the bond angles is also significant (109.8° for the bulk silicon): for the PP parts the angles range from 105.0° to 112.3° and for the IQD part from 107.2 ° to 110.2°.

To study the SiNF electronic structure we performed the Extended Huckel Theory (EHT)[21] calculations of the SiPNW/D (205 atoms, this structure was obtained by truncation from the SiNF petal), the SiNW/H ($Si_{205}H_{150}$) and SiNW/HOOH ($Si_{265}(OH)_{20}O_{185}$). All molecular diagrams presented on the Fig. 3 were obtained for geometry, optimized using MM+ model potential.

The comparison of the EHT calculations with the DFT [9, 11-13] ones shows that the EHT method correctly describes the nature of the electronic system, displaying the metallic properties of the pristine structures and semiconducting gaps of the saturated systems (Table 1). The EHT calculations systematically overestimate (0.8 – 2 eV) the band gap of all semiconducting systems excluding the $SiIQD_{600}$/H structure (overestimation 4.045 eV).

Like the high quality DFT calculations, [9, 11-13] the EHT level of theory (Table 1, Fig. 3) shows a metallic state for the pristine silicon nanoflower (Fig. 3a). The hydrogenated nanoflower reveals a wide band gap (6.8 eV for SiNF/H, Fig. 3b). The average atomic charge of silicon atoms for the SiNF/H and SiNW/H systems at EHT level of theory is close to +0.35.

The EHT calculations of the SiNF/OOH (Fig. 3c) and SiNW/OOH systems give significantly lower values for the band gap (2.0 and 2.3 eV correspondingly). These values correspond well to the B3LYP/6-31G* calculations (1.5 eV) [11] of the SiNW/O system with the same length and oxidize layer. The average silicon EHT atomic charge is close to +1.3, whereas the B3LYP/6-31G* ones [11] are lower and close to +1.0.

The EHT calculation of the triple SiNF system (Fig. 3d) gives 0 band gap. Both HOMO and LUMO states are localized at the silicon petal uncovered by hydrogen or oxygen. Two different semiconducting petals keep the same charge distribution as the parent SiNW/H and SiNW/OOH systems. So, the triple system, with 3 different PP types (one metallic and two semiconducting), can act as a structural unit of nanoelectronic devices.

The combination of the IQD core with the PP parts produce other promising X, Y and V planar structures. Each IQD has six five-fold symmetry axes through vertexes and each pair of the axes belongs to one symmetry plane. The planar nanowire junctions through the IQDs can be formed with four (X), three (Y) and two (V) SiNWs. In the last case the two angles ($60^o$ and $120^o$) between the petals can be realized by truncation of different terminals from the Y-structure. It is necessary to note that the two-terminal linear structure can be formed by truncation of corresponding petals from Y- or X-structures. It has a $D_{5d}$ point symmetry group with two PPs rotated around each other at a $36^o$ angle. The two- or three-terminal junctions can serve as nanodiods and nanotransistors or as units of logic networks. [22] The proposed SiNW/SiIQD structures can be used as parts of nanomechanical devices and micromachines or as a filling agent for increasing the strength of different composites due to their structural features.

We have presented a new class of silicon nanostructures based on icosahedral central core and different number of pentagonal petals from 2 up to 12 growing from some or all vertexes of the central quantum dot. The formation of the SiNW/SiIQD interface is energetically preferable due to decreasing of the total surface tension of the system. Some observed exotic snowflake silicon micro-[16] and nano- [17] structures could be formed on the base of considered SiNFs as precursors. The unique physical properties of the proposed nanostructures make the SiNFs and SiNW/SiIQD junctions promising candidates for wide variety of applications as structural units of nanoelectronic and nanomechanical devices.

ACKNOWLEDGMENTS

This work was partially supported by project "Materials Design with New Functions Employing Energetic Beams" and JAEA Research fellowship (PVA), by Russian Fund of Basic Researches (grant 05-02-17443), grant of Deutsche Forschungsgemeinschaft and Russian Academy of Sciences, No. 436 RUS 113/785 (LAC) and from the US Department of Energy via the Ames Laboratory and the Air Force Office of Scientific Research. Partially, the calculations have been performed on the Joint Supercomputer Center of the Russian Academy of Sciences. The geometry of all presented structures was visualized by ChemCraft software. [23] PVA also acknowledges the members of "Research Group for Atomic-scale Control for Novel Materials under Extreme Conditions" Prof. H.

FIGURE CAPTIONS

Fig. 1.a) Perpendicular cross section of the SiNW. The SiNW(1) part is marked by vinous. b) The atomic structure of IQD/PP interface (IQD is presented partially). The silicon atoms of the icosahedral core forming the pentagon vertexes are presented in pink. The PP atoms forming the IQD/PP interface are presented in orange. Blue arrows represent the chemical bonds between IQD and PP. c) The relative stability (eV/atom, MM+ level of theory) of the different types of silicon nanostructures (dark and light circles are the SiNW and IQD energies respectively, triangle is the IQD/PP energy) in respect with the MM+ binding energy of the bulk silicon (0.0 eV/atom).

Fig. 2. a) SiNF with 10 petals covered by oxidized layers (oxygen atoms are in red and silicon atoms are in pink) and the stalk covered by hydrogen atoms (in blue). The petals and the stalk are attached to the SiIQD1100. b) The SiNF with 10 petals covered by an oxidized layer. The silicon atoms are in vinous and oxygen atoms are in green. c) Silicon nanoflower with 5 petals.

Fig. 3. A set of 3-petalled silicon nanoflowers with different types of petals and the corresponding molecular diagrams calculated at the EHT level of theory. The occupied levels are presented in blue and vacant levels are presented in beige. a) Left: pristine silicon SiNF/D structure. The (100) surfaces of all petals are relaxed to form the surface dimers (D). Right: molecular diagrams of the SiIQD$_{100}$ and SiNF/D. b) Left: the SiNF/H structure. All dangling bonds are saturated by hydrogen (in blue). Right: molecular diagrams of SiIQD$_{100}$/H, SiNW$_{205}$/H and SiNF/H. c) Left: the SiNF/OOH structure. All dangling bonds are saturated by bridged oxygen, SiO$_4$ and OH groups (the oxygen atoms are presented in green). Right: molecular diagrams of SiNW$_{205}$/OOH and SiNF/OOH. d) Left: combined SiNF/D/H/OOH structure. Right: molecular diagrams of the SiNF/D/H/OOH and SiNW$_{205}$ structures.

Table 1. The HOMO-LUMO gap for different semiconducting SiNWs and SiNFs, eV

|  | EHT | DFT |
|---|---|---|
| SiIQD$_{100}$ | 0 | -- |
| SiIQD$_{100}$/H | 7.633 | -- |
| SiIQD$_{600}$/H | 5.775 | 1.73 [12] |
| Si$_{205}$NW | 0 | 0 [9-11, 13] |
| Si$_{205}$NW/H | 7.294 | 5-6, depending the size of the NWs [15] |
| Si$_{205}$NW/OOH | 2.296 | 1.5 [13] |
| SiNF/H | 6.787 | -- |
| SiNF/OOH | 1.939 | -- |
| SiNF | 0 | -- |
| Combined SiNF/H/OOH | 0 | -- |

Fig 1.

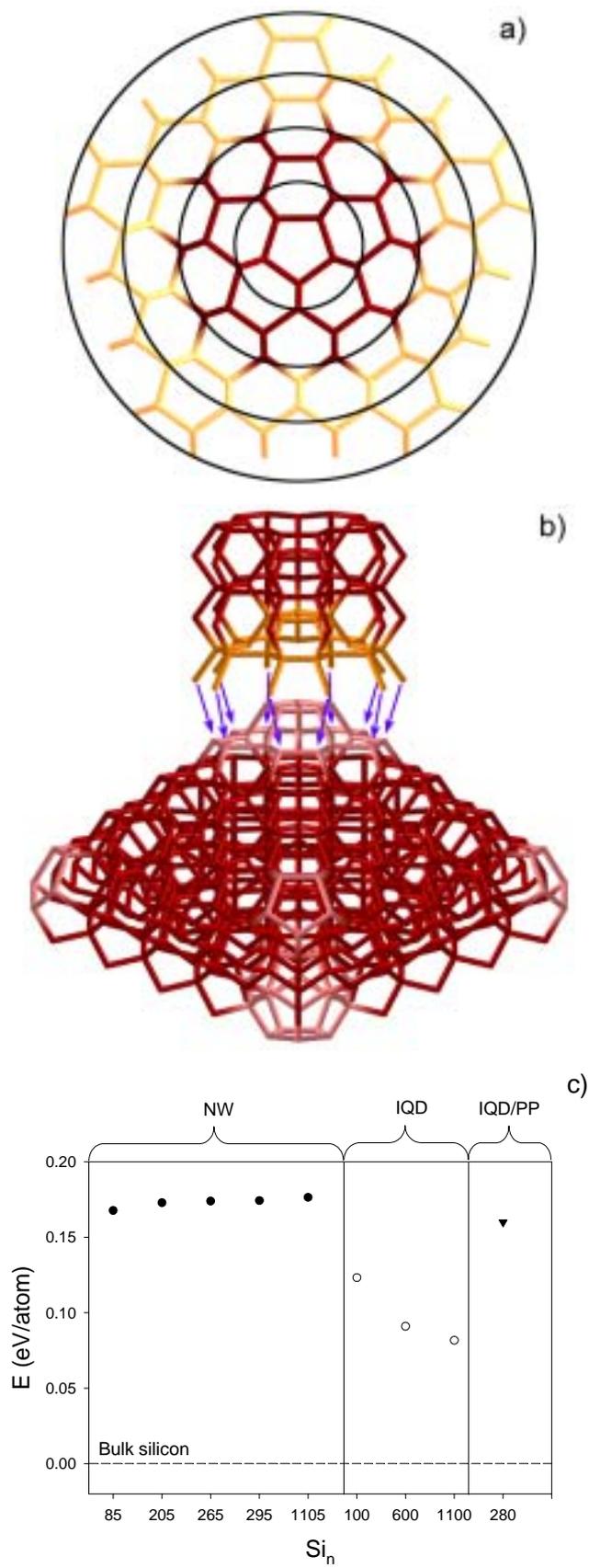

Fig. 2

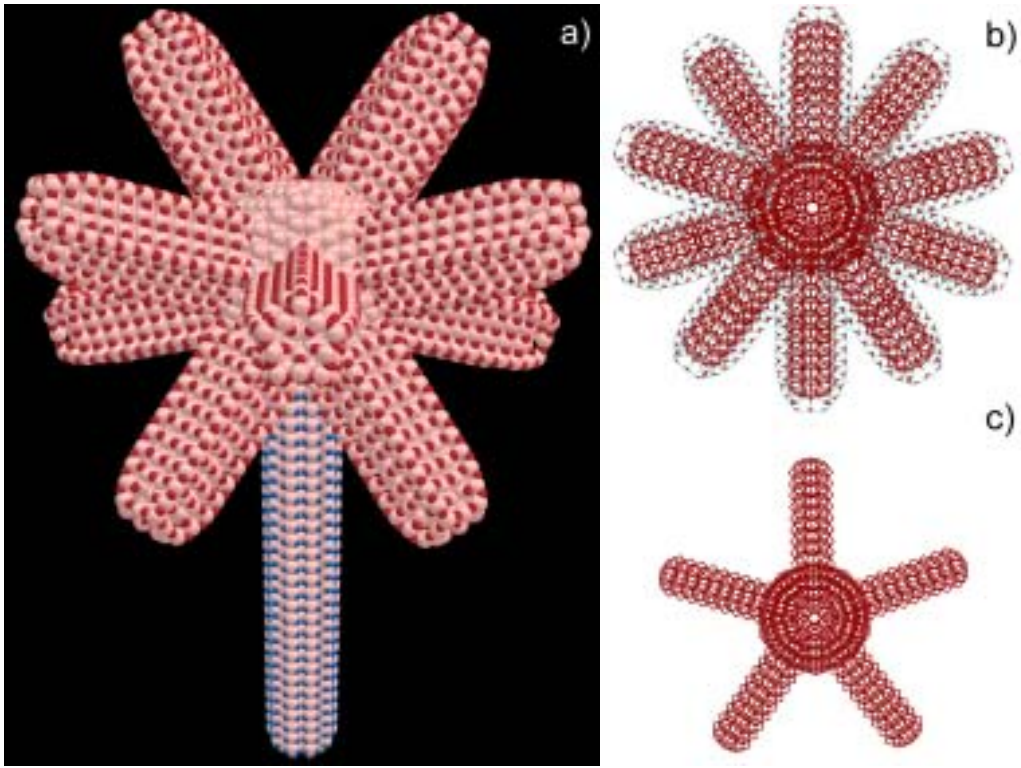

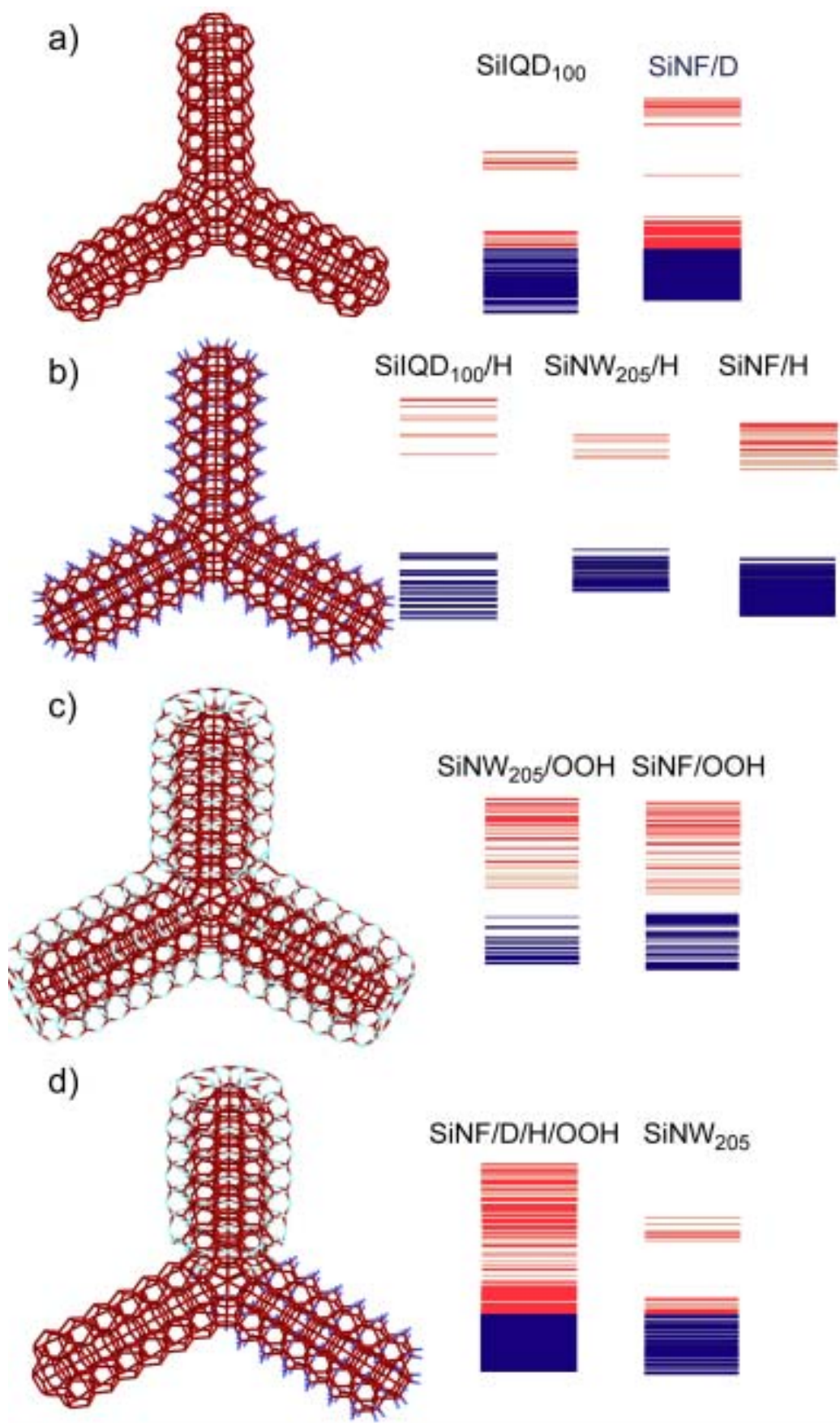